\documentclass[prc,11pt, amsmath, amsymb, 
aps,reprint,tightenlines,nofootinbib,longbibliography,abbrv,preprintnumbers]{revtex4-1}
\usepackage[utf8]{inputenc}
\usepackage{graphicx}   
\usepackage[colorlinks=true,linkcolor=blue,citecolor=blue, urlcolor=blue]{hyperref}   
\usepackage{bm} 

\def\p{{\bf p}}

\def\st{\begin{equation}}
\def\stp{\end{equation}}
\def\bg{\begin{eqnarray}}
\def\nd{\end{eqnarray}}
\def\Eq#1{Eq.~(\ref{#1})}
\def\Eqs#1{Eqs.~(\ref{#1})}
\def\eq#1{(\ref{#1})}
\def\app#1{Appendix~\ref{#1}}
\def\Fig#1{Fig.~\ref{#1}}

\def\Sect#1{Sect.~\ref{#1}}
\def\Sec#1{Sec.~\ref{#1}}
\def\Ref#1{Ref.~\cite{#1}}

\def\llangle{\left\langle}
\def\rrangle{\right\rangle}

\def\t22{{\scriptscriptstyle T_2T_2}}

\def\x{{\bm x}}

\def\k{{\bm k}}

\begin{document}

\title{Bulk viscosity from hydrodynamic fluctuations with relativistic hydro-kinetic theory}
\author{Yukinao Akamatsu}
\email{akamatsu@kern.phys.sci.osaka-u.ac.jp}
\affiliation{Department of Physics, Osaka University, Toyonaka, Osaka 560-0043, Japan}
\author{Aleksas Mazeliauskas}
\email[]{a.mazeliauskas@thphys.uni-heidelberg.de}
\affiliation{Institut f\"{u}r Theoretische Physik, Universit\"{a}t Heidelberg, 
	D-69120 Heidelberg, Germany}
\affiliation{Department of Physics and Astronomy, Stony Brook University, Stony Brook, New York 11794, USA}
\author{Derek Teaney}
\email[]{derek.teaney@stonybrook.edu}
\affiliation{Department of Physics and Astronomy, Stony Brook University, Stony Brook, New York 11794, USA}

\date{\today}

\begin{abstract}
Hydrokinetic theory of thermal fluctuations is applied to a nonconformal relativistic fluid.
Solving the hydrokinetic equations for an isotropically expanding background we find that hydrodynamic fluctuations give ultraviolet divergent contributions to the energy-momentum tensor.
After shifting the temperature to account for the energy of nonequilibrium modes, the remaining divergences are renormalized into local parameters, e.g.,\ pressure and bulk viscosity. 
We also confirm that the renormalization of the pressure and bulk viscosity is universal by computing them for a Bjorken expansion.
The fluctuation-induced bulk viscosity reflects the nonconformal nature of the equation of state and is modestly enhanced near the QCD deconfinement temperature.
\end{abstract}


\maketitle


\section{Introduction}
Ultrarelativistic heavy-ion collisions  are a major experimental tool to study nuclear matter in an extremely hot environment. The energy density in heavy ion collisions at the Relativistic Heavy Ion Collider (RHIC) at BNL and the Large Hadron Collider (LHC) at CERN is so high that partonic degrees of freedom are liberated from nucleons and a deconfined quark-gluon plasma (QGP) is formed.
The QGP then expands hydrodynamically as a fluid with very small shear viscosity over entropy ratio $\eta /s =(1$--$2)/(4\pi)$~\cite{Heinz:2013th,Luzum:2013yya}. 
The hydrodynamic paradigm for heavy-ion collisions has been very successful in explaining the various collective flow observables as a dynamical response to event-by-event fluctuations of the initial fireball shape~\cite{Heinz:2013th,Teaney:2009qa,Luzum:2013yya,Gale:2013da,Romatschke:2009im}.

Recently, attention has been paid to another source of fluctuations in the hydrodynamic picture, namely, thermal fluctuations~\cite{Gavin:2006xd,Kapusta:2011gt,Yan:2015lfa,Young:2014pka,Kapusta:2012zb,Murase:2016rhl,Nagai:2016wyx}.
Thermal fluctuations are theoretically required by the fluctuation-dissipation theorem.
Furthermore, thermal fluctuations play an important role in systems with a small number of particles and are essential near the critical point, which is the focus of the ongoing beam energy scan program at RHIC~\cite{Kumar:2013cqa}.

A unique feature of hydrodynamic fluctuations in heavy-ion collisions is the rapidly expanding background flow along the beam direction, which at midrapidity is often modelled as one-dimensional Bjorken flow~\cite{Bjorken:1982qr}.
The distribution of fluctuations around such evolving background is characterized  by a specific wave number scale $k_*$, where the longitudinal expansion and ($k$-dependent) relaxation rates balance, and the distribution function approaches a nonequilibrium steady state.
In the previous publication, we developed an effective kinetic description for \emph{conformal} hydrodynamic fluctuations around the characteristic scale $k_*$ and discussed how to deal with ultraviolet divergences associated with short
wavelength fluctuations~\cite{Akamatsu:2016llw}.
Using the hydrokinetic theory we obtained a universal renormalization of the pressure and shear viscosity in agreement with previous diagrammatic calculations around a nonexpanding background~\cite{Kovtun:2003vj,Kovtun:2011np}. 
Furthermore, we applied the hydrokinetic approach to the Bjorken expansion, and found the precise coefficient of the fractional-power-law tail arising from the out-of-equilibrium distribution of hydrodynamic fluctuations.

In this paper, we consider a relativistic \emph{nonconformal} fluid, for which the speed of sound $c_s^2(T)\neq 1/3$ and the bulk viscosity is finite.
The bulk viscosity determines the dissipative correction to the pressure in response to an isotropic expansion or compression and is a measure for scale symmetry breaking.
For example, perturbative calculations in a high-temperature QGP show that it is proportional to the square of the scale symmetry breaking factors (the QCD running coupling and finite quark mass)~\cite{Arnold:2006fz}.
Also, lattice QCD simulations suggest a correlation between the bulk viscosity and the scale symmetry breaking realized in the equation of state~\cite{Meyer:2007dy}.
Spectral sum rules in the bulk channel also indicate some correlation between the bulk viscosity and a nonconformal nature of the equation of state~\cite{Kharzeev:2007wb,Karsch:2007jc,Moore:2008ws,Romatschke:2009ng}.
Finally, near the critical point, the bulk viscosity diverges because of the critical slowing down~\cite{Onuki}.

In the main part of the paper we apply our hydrokinetic theory to a static system perturbed by an isotropic expansion and compute the response function of the energy-momentum tensor in the bulk channel.
We discuss the case of Bjorken expansion in \app{app:bjorken}.
In a nonconformal fluid the two-point correlation function of hydrodynamic fluctuations contributes to the trace of the energy momentum tensor, which gives rise to a renormalization of the bulk viscosity:
\begin{align}
\label{eq:zetaIntr}
\zeta(T)&=
\zeta_0(T;\Lambda) \\
& \quad+\frac{T\Lambda}{18\pi^2}
\left[
\begin{aligned}
&\left(1+\frac{3T}{2}\frac{dc_{s0}^2}{dT} -3c_{s0}^2\right)^2\frac{e_0 + p_0}{\zeta_0+\frac{4}{3}\eta_0} \\
&+4\left(1-3c_{s0}^2\right)^2\frac{e_0 + p_0}{2\eta_0}
\end{aligned}
\right]. \nonumber
\end{align}
Here, $\Lambda$ is a UV cut-off for the hydrodynamic fluctuations and $\zeta_0(T;\Lambda)$ is the bare bulk viscosity.
The fluctuation contribution to the bulk viscosity is positive and proportional to the scale symmetry breaking factors in the equation of state. 
It is noteworthy that to arrive at \Eq{eq:zetaIntr}, the temperature of the background fluid must be shifted depending on the cut-off so as to include the energy of the non-equilibrium hydrodynamic modes (see \Sec{sec:tempshift} for details).

\begin{figure}
	\centering
	\includegraphics[width=\linewidth]{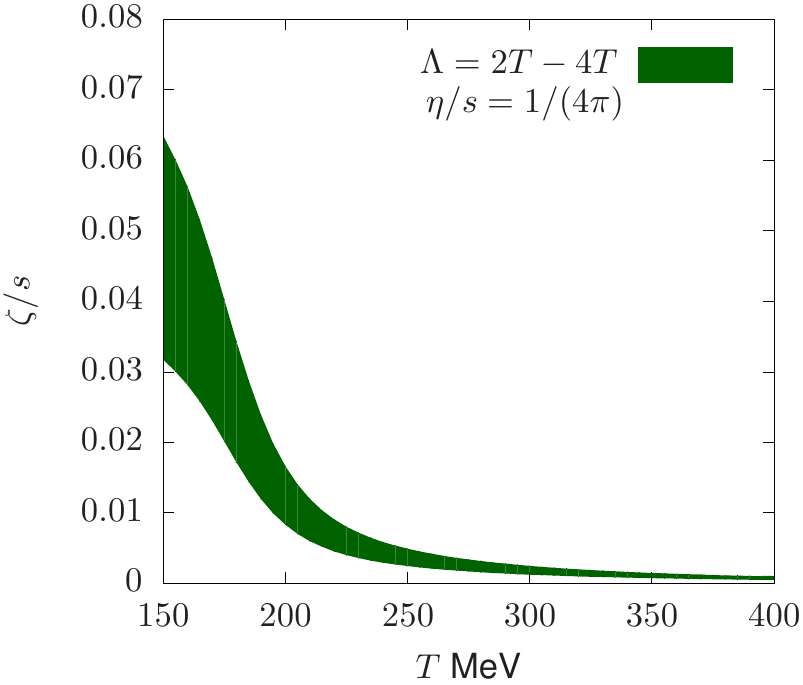}
	\caption{A fluctuation-induced bulk viscosity bound as a function of temperature, \Eq{eq:zetaIntr}, for lattice parametrization of the QCD equation of state and shear viscosity over entropy $\eta/s=1/(4\pi)$~\cite{Borsanyi:2016ksw}.
	The UV bound  of hydrodynamic fluctuations $\Lambda$ is varied between $2T$ and $4T$.}
	\label{fig:bulk}
\end{figure}

The fluctuation-induced renormalization in \Eq{eq:zetaIntr} can be used to estimate a lower bound of the bulk viscosity of QCD --- see \Ref{Kovtun:2011np} for a  similar estimate of the shear viscosity.
Very recently the approach  was also used to estimate the bulk viscosity of a nonrelativistic cold Fermi gas, where the
renormalization was obtained with  diagrammatic methods~\cite{Martinez:2017jjf} (we performed the diagrammatic calculation for the relativistic nonconformal fluid in \app{app:diagrams}).
Using the lattice equation of state for entropy density $s(T)$ and  the speed of sound $c_s^2(T)$~\cite{Borsanyi:2016ksw} in \Eq{eq:zetaIntr}, we calculate the magnitude of bulk viscosity renormalization by setting $\zeta_0=0$, and choosing representative values of the kinematic viscosity, $\eta/s=1/4\pi$, and the temperature-dependent UV
cut-off $\Lambda=2T-4T$ (see \Fig{fig:bulk}). 
Because of the small deviation from scale symmetry at high temperatures the bulk viscosity renormalization is vanishing small for $T\gg T_c$.
However, the degree of nonconformality $(c_s^2-\frac{1}{3})^2$ peaks around the pseudocritical temperature where the bulk viscosity reaches $\zeta/s\sim0.03-0.06$ at $T_c\sim150\,\text{MeV}$. 

The logic of the estimate in \Fig{fig:bulk} is the following.
The physical bulk viscosity $\zeta(T)$ (which is independent of $\Lambda$) arises from two contributions: the fluctuations above $\Lambda$, which at weak coupling are dominated by single-particle excitations, and the fluctuations below $\Lambda$,  which are described by hydrodynamics.
We have only included the hydrodynamic fluctuations here, and thus we expect the physical bulk viscosity to be larger than the estimate shown in \Fig{fig:bulk}.  

The organization of this paper is as follows.
In Sec.~\ref{sec:kinetics}, we derive the kinetic equations for hydrodynamic fluctuations for an isotropically expanding nonconformal fluid.
Then in Sec.~\ref{sec:nonlinear}, we compute the fluctuation contributions to the energy-momentum tensor, and discuss the subtle temperature shift.
After the temperature shift, we renormalize the energy density, the pressure, and the bulk viscosity, and find the finite long-time tails for the weak isotropic expansion.
The summary of the paper is given in Sec.~\ref{sec:summary}. Finally, in  \app{app:bjorken} we repeat the computation of the temperature shift and the renormalization of hydrodynamic fields for Bjorken expansion.
In \app{app:diagrams}, we give a diagrammatic derivation for the bulk viscosity renormalization, which is consistent with our results by the hydrokinetic theory.


\section{Kinetic equations for hydrodynamic fluctuations}
\label{sec:kinetics}
In this section we apply the formalism developed in \Ref{Akamatsu:2016llw} to a nonconformal fluid under isotropic expansion (or compression).
We will follow the same procedure to derive the relaxation type equations for the two-point correlation functions under the presence of background perturbations.

The governing equations for nonconformal hydrodynamics with noise are given by~\cite{LandauStatPart1, LandauStatPart2, [for a recent review: ]Kovtun:2012rj}
\begin{subequations}
\label{eq:hydro+noise}
\begin{align}
d_\mu T^{\mu\nu} &=0\label{eomzero},\quad 
T^{\mu\nu}=T^{\mu\nu}_\text{ideal}+T^{\mu\nu}_\text{visc.}+S^{\mu\nu},\\
T^{\mu\nu}_{\rm ideal} &= (e + p) u^{\mu} u^{\nu} + pg^{\mu\nu},\\
T^{\mu\nu}_{\rm visc.} &= - \eta \sigma^{\mu\nu} - \zeta \Delta^{\mu\nu}\Delta^{\alpha\beta}d_{\alpha}u_{\beta},\\
\sigma^{\mu\nu}  &= \Delta^{\mu\rho}\Delta^{\nu\sigma} (d_{\rho} u_{\sigma} + d_{\sigma} u_{\rho} - \frac{2}{3}
g_{\rho\sigma} d_{\gamma} u^{\gamma}), \\
\Delta^{\mu\nu} &= g^{\mu\nu} + u^{\mu} u^{\nu},
\end{align}
\end{subequations}
where $d_{\mu}$ denotes a covariant derivative using the ``mostly-plus" metric convention.
Below we notate the divergence of the flow velocity as $\nabla \cdot u \equiv d_{\mu} u^{\mu}$.
The variance of the stochastic noise is determined by the fluctuation-dissipation theorem:
\begin{align}
&\langle S^{\mu\nu}(x_1)S^{\alpha\beta}(x_2)\rangle \nonumber\\
&= 2T\left[
\begin{aligned}
&\eta \left(\Delta^{\mu\alpha}\Delta^{\nu\beta} + \Delta^{\mu\beta}\Delta^{\nu\alpha}\right) \\
&+\left(\zeta - \frac{2}{3}\eta\right)\Delta^{\mu\nu}\Delta^{\alpha\beta}
\end{aligned}
\right] \frac{\delta(x_1-x_2)}{\sqrt{-\det g_{\mu\nu}}} .
\end{align}
Differently from the conformal case, both shear $\eta$ and bulk $\zeta$ viscosities are now present in the equation of motion and noise correlator.
\subsection{Background fluid}
Dynamics of hydrodynamic fluctuations on a background fluid in a weak isotropic expansion (or compression) is conveniently studied in the reference frame of the fluid.
In the comoving frame for the isotropic expansion, the metric is time dependent,
\begin{align}
ds^2=-dt^2 + (1+h(t))d\vec x^2, \ \ \ (|h(t)|\ll 1)
\end{align}
and the background fluid satisfies
\begin{align}
\label{eq:hydroeq_bkg}
0=\dot e_0(t) + \frac{3\dot h}{2}[e_0(t) + p_0(t)] + \mathcal O(h^2).
\end{align}
The second term on the right-hand side represents the change of energy density from the expansion and the associated work done by the pressure.
Throughout this paper, $X_0$ denotes a quantity $X$ of the background fluid in a perturbed metric ($h\neq 0$).
As discussed previously~\cite{Akamatsu:2016llw}, $e_0(t)$ and $p_0(t)$ denote the background  energy density and pressure from modes with wavenumbers greater than a cut-off $\Lambda$.
In \Sec{sec:tempshift} we detail how $e_0$ and $p_0$ are related to the lattice equation of state.

Solving perturbatively in $h$, the energy density $e_0(t)$ for the background fluid evolves as
\begin{align}
\label{eq:e_bkg}
e_0(t) = \bar e_0 -\frac{3h(t)}{2}(\bar e_0+\bar p_0) + \mathcal O(h^2),
\end{align}
where $\bar e_0$ denotes the energy density of the background fluid in an unperturbed state ($h=0$).
Again, throughout this paper $\bar{X}_0$ denotes a quantity $X$ of the background fluid in an unperturbed state ($h=0$).

\subsection{Evolution of hydrodynamic fluctuations}

For the expanding background described by \Eq{eq:e_bkg}, the hydrodynamic fluctuations excited by thermal noise $\delta e(t,\x)\equiv e(t,\x)-e_0(t)$ and $\vec{g}\equiv(e_0(t)+p_0(t))\vec{v}(t,\x)$ evolve according to the following equations in $\k$ space:
\begin{subequations}
\label{eq:eom1}
\begin{align}
0&=\partial_t\delta e + ik^i g_i + \frac{3\dot h}{2}(1+c_{s0}^2)\delta e,\\
0&=\partial_t g_i + i c_{s0}^2 k_i \delta e + \frac{3\dot h}{2}g_i \nonumber \\
&\quad +\gamma_{\eta 0} (k^l k_l \delta_i^j-k_ik^j) g_j + \gamma_{\zeta0}k_ik^j g_j+\xi_i,
\end{align}
\end{subequations}
with noise correlation given by
\begin{align}
\langle \xi_i(t, \bm k) \xi_j(t', -\bm k')\rangle &= \frac{2T_0(e_0+p_0)}{\sqrt{-\text{det}\,g_{\mu\nu}}}(2\pi)^3\delta(\bm k-\bm k')\delta(t-t')\nonumber \\
&\quad \times\left[\gamma_{\eta 0} (k^l k_l g_{ij}-k_ik_j) + \gamma_{\zeta 0}k_ik_j\right].
\end{align}
Here $\gamma_{\eta}\equiv \eta/(e+p)$ and $\gamma_{\zeta}\equiv (\zeta+\frac{4}{3}\eta)/(e+p)$ are kinematic viscosities.
Analysis becomes simpler by utilizing a vielbein formalism.
We introduce new variables
\begin{subequations}
\begin{align}
G_{\hat i} &\equiv \left(1+\frac{1}{2}h(t)\right)g^i, \\
K_{\hat i} &\equiv \left(1-\frac{1}{2}h(t)\right)k_i,\\
\Xi_{\hat i} & \equiv \left(1+\frac{1}{2}h(t)\right)\xi^i,
\end{align}
\end{subequations}
which give $G_{\hat i}G_{\hat i}=g_i g^i$, $K_{\hat i}K_{\hat i}=k_i k^i$, and $G_{\hat i}K_{\hat i} = g^i k_i=g_i k^i$.
We define a four-component vector $\phi_a\equiv (c_{s0}\delta e,\vec G)$ of hydrodynamic fluctuations.
The equation of motion for $\phi_a$ is
\begin{subequations}
   \label{eq:matrices}
\begin{align}
\label{eq:phi_evol}
-\dot\phi_a(t,\bm k)
&=i\mathcal{L}_{ab}\phi_b + \mathcal{D}_{ab}\phi_b + \Xi_a + \mathcal{P}_{ab}\phi_b,\\
\label{eq:matrixL}
&\hspace{-4em}\mathcal{L} = \left(
\begin{array}{cc}
0 & c_{s0}\vec K\\
c_{s0}\vec K & 0
\end{array}
\right), \\
\label{eq:matrixD}
&\hspace{-4em}\mathcal{D} = \left(
\begin{array}{cc}
0 & 0\\
0 & 
\gamma_{\eta 0}\left(K^2\delta_{\hat i\hat j}-K_{\hat i}K_{\hat j}\right)
+ \gamma_{\zeta 0}K_{\hat i}K_{\hat j}
\end{array}
\right), \\
\label{eq:matrixP}
&\hspace{-4em}\mathcal{P} = \dot h\left(
\begin{array}{cccc}
\frac{3}{2}\left(1+\bar c_{s0}^2 + \frac{\bar T_0}{2}\frac{d\bar c_{s0}^2}{d\bar T_0}\right) & & &\\
 & 2 & & \\
 & & 2 & \\
 & & & 2
\end{array}
\right),
\end{align}
\end{subequations}
with noise correlation given by
\begin{align}
&\langle \Xi_a(t,\bm k)\Xi_b(t', -\bm k') \rangle \nonumber \\
&\quad =\frac{2T_0(e_0+p_0)}{\sqrt{-\text{det}\,g_{\mu\nu}}}\mathcal{D}_{ab} (2\pi)^3\delta(\bm k-\bm k')\delta (t-t').
\end{align}
The matrices $\mathcal L$ and $\mathcal D$ originate from ideal and viscous parts of the hydrodynamic equations respectively, while $\mathcal P$ arises from the remaining interactions between the fluctuations and the background fluid.
Note that the term $\propto \frac{\bar T_0}{2}\frac{d \bar c_{s0}^2}{dT_0}$ in $\mathcal P$ derives from the time dependence of $c_{s0}(T_0(t))\delta e$ in $\phi_1$.  
In the kinetic regime, $\mathcal{L}$ drives the evolution of $\phi_a$ so that it will be more convenient to analyze \Eq{eq:phi_evol} in terms of eigenmodes of $\mathcal L$:
\begin{align}
(e_{\pm})_a = \frac{1}{\sqrt{2}}
\begin{pmatrix}
1 \\
\pm\hat{K}
\end{pmatrix},
\quad
(e_{T_1})_a =
\begin{pmatrix}
0 \\
\vec{T}_1 \\
\end{pmatrix},
\quad
(e_{T_2})_a =
\begin{pmatrix}
0 \\
\vec{T}_2 \\
\end{pmatrix}.\nonumber \\
\label{eq:vectors}
\end{align}
Here $\hat K \equiv \vec K/|K|$, $\vec T_1$, and $\vec T_2$ form an orthonormal basis.
The subscripts $+,-$  stand for the two sound modes and $T_1,T_2$  for the two transverse diffusive modes.
The corresponding eigenvalues are $\lambda_{\pm}=\pm c_s K$ and $\lambda_{T_1,T_2}=0$.

\subsection{Kinetic equations for hydrodynamic fluctuations}
The two-point correlation functions of $\phi_{A} \equiv \phi_a \left(e_{A}\right)_a$ with $A=+,-,T_1, T_2$ are defined as
\begin{align}\label{eq:NAA}
\llangle \phi_A(t,\bm k) \phi_B(t,-\bm k')\rrangle \equiv N_{AB}(t,\bm k)(2\pi)^3\delta(\bm k - \bm k') .
\end{align}
We will determine the equations of motion for $N_{AB}(t,\k)$ using the formalism of \Ref{Akamatsu:2016llw}.
In the rotating wave approximation, the off-diagonal part of the density matrix $N_{AB}$ can be neglected because of its rapid phase rotation\footnote{
$N_{T_1T_2}$ has a stationary phase but vanishes because of the rotational symmetry.
}, while the diagonal part evolves according to
\begin{align}
\label{eq:2pt-func}
\dot N_{AA} = -2\mathcal{D}_{AA}\left[N_{AA} - \frac{T_0(e_0+p_0)}{\sqrt{-\det g_{\mu\nu}}}\right]
+2\mathcal{P}_{AA}N_{AA},\nonumber \\
\end{align}
where we have defined $\mathcal{D}_{AA}\equiv (e_{A})_a \mathcal{D}_{ab}(e_A)_b$ and similarly $\mathcal{P}_{AA}$.
The isotropic system does not distinguish the two transverse modes $T_1$ and $T_2$, and thus we only have two independent kinetic equations: one  for the sound modes ($L=++,--$), and one for the transverse modes ($T=T_1T_1,T_2T_2$). 
Using the matrices and eigenvectors of the previous section, \Eq{eq:2pt-func} evaluates to
\begin{subequations}
\begin{align}
\label{eq:hydrokin_L}
\dot N_{L} &= -\gamma_{\zeta 0} K^2 \left[N_{L} - \frac{T_0(e_0+p_0)}{\sqrt{-\det g_{\mu\nu}}}\right] \nonumber \\
& \quad -\frac{\dot h}{2}\left(3\bar c_{s0}^2+\frac{3\bar T_0}{2}\frac{d\bar c_{s0}^2}{d\bar T_0}+7\right)N_{L},\\
\label{eq:hydrokin_T}
\dot N_{T} &= -2\gamma_{\eta 0} K^2 \left[N_{T}- \frac{T_0(e_0+p_0)}{\sqrt{-\det g_{\mu\nu}}}\right]-4\dot h N_{T}.
\end{align}
\end{subequations}
The kinetic equations \eqref{eq:hydrokin_L} and \eqref{eq:hydrokin_T} describe how the distribution of fluctuations $\phi_A$ evolves on the isotropically expanding background.
Perturbative solutions of the kinetic equations for $|h|\ll 1$ take the form,
\begin{align}
\label{eq:kinsol}
N_{L/T}(t,\bm k)&=N_{\rm eq}(t) + \delta N_{L/T}(t,\bm k) + \mathcal O(h^2), 
\end{align}
where the equilibrium contribution is
\begin{align}
\label{eq:kinsol_eq}
N_{\rm eq}(t)&=\frac{T_0(e_0+p_0)}{\sqrt{-\det g_{\mu\nu}}} \nonumber \\
&\simeq \left[1-(3+3\bar c_{s0}^2)h(t)\right]\bar T_0(\bar e_0 + \bar p_0),
\end{align}
and the non-equilibrium correction $\delta N_{L/T}$ is
\begin{subequations}
   \label{eq:kinsol_neq}
\begin{align}
\label{eq:kinsol_L}
\delta N_L(\omega,\bm k)&=\frac{\frac{1}{2}i\omega h(\omega)}{-i\omega +\bar\gamma_{\zeta 0} K^2}
\bar C_{\zeta 0} \bar T_0(\bar e_0+\bar p_0),\\
\label{eq:kinsol_T}
\delta N_T(\omega,\bm k)&=\frac{i\omega h(\omega)}{-i\omega + 2\bar\gamma_{\eta 0} K^2}
\bar C_{\eta 0}\bar T_0(\bar e_0+\bar p_0).
\end{align}
\end{subequations}
Here and below we have defined
\begin{subequations}
	\label{eq:C}
\begin{align}
C_{\zeta}(T) &\equiv  1+\frac{3T}{2}\frac{dc_s^2}{dT}-3c_s^2,\\
C_{\eta}(T) &\equiv  1-3c_s^2.
\end{align}
\end{subequations}
Note that when the background fluid is scale invariant $e_0=3p_0$, the corrections $\delta N_{L/T}$ vanish.
Therefore in conformal case, the isotropic expansion or compression does not drive the hydrodynamic fluctuations from the equilibrium distribution $N_\text{eq}(t)$ given by \Eq{eq:kinsol_eq}. 

For $k\sim k_{*}$ the distribution of fluctuations in \Eq{eq:kinsol_neq} is not well characterized by the time derivatives of $h(t)$.
However, at large $k$ by the distribution approaches equilibrium with calculable first derivative corrections:\footnote{In the current setup  $\nabla \cdot u=\tfrac{3}{2} \dot h$. }
\begin{subequations}
   \label{eq:nhighk}
   \begin{align}
\delta N_L(t,\bm k)&\simeq - \frac{3
\bar C_{\zeta 0} \bar T_0(\bar e_0+\bar p_0)  
      }{\bar\gamma_{\zeta 0} K^2} \, 
      \nabla \cdot u,\\
\label{eq:kinsol_T}
\delta N_T(t,\bm k)&\simeq -\frac{3
\bar C_{\eta 0}\bar T_0(\bar e_0+\bar p_0)  
      }{2\bar\gamma_{\eta 0} K^2} \, 
       \nabla \cdot u.
\end{align}
\end{subequations}
It is these corrections $\propto\nabla\cdot u/K^2$ which are responsible for the renormalization of the bulk viscosity  and the
temperature shift described in \Sect{sec:nonlinear}.


\section{Energy-momentum tensor with nonlinear fluctuations}
\label{sec:nonlinear}
In this section we compute the nonlinear contributions of hydrodynamic fluctuations to the statistically averaged energy momentum tensor $\left<T^{\mu\nu}\right>$.
The main difference from the conformal case~\cite{Akamatsu:2016llw} is additional contributions to the averaged energy
density $\left<T^{tt}\right>$, which are absorbed by a shift in the background temperature $T_0(t,\Lambda)$. 
\subsection{Averaged energy-momentum tensor}
The averaged stress tensor consists of contributions from the background fluid and from the two-point functions of the hydrodynamic fluctuations:
\begin{subequations}
\begin{align}
   \label{Tijtotdef}
\langle T^{ij}\rangle&=\left[1-h(t)\right]p_0\delta^{ij} - \frac{3}{2}\dot h(t)\zeta_0\delta^{ij} + T^{ij}_{\rm fluct},\\
T^{ij}_{\rm fluct}
&\simeq \frac{1-h(t)}{e_0+p_0}\left[
\begin{aligned}
&\langle G_{\hat i}(t,\bm x) G_{\hat j}(t,\bm x)\rangle  \\
&+ \delta^{ij}\frac{T_0}{2}\frac{dc_{s0}^2}{dT_0}\langle (c_{s0}\delta e(t,\bm x))^2\rangle
\end{aligned}
\right]. 
\end{align}
\end{subequations}
The energy density fluctuations $\propto \langle (c_{s0}\delta e(t,\bm x))^2\rangle$ originate from the second-order derivative $\frac{d^2 p_0}{de_0^2}$, which is finite for a nonconformal equation of state.
The trace of the stress tensor from the fluctuations is determined by the two-point functions $N_{L/T}$:
\begin{align}
T^{ii}_{\rm fluct}
&= \frac{1-h(t)}{e_0+p_0}\int\frac{d^3k}{(2\pi)^3}\\
& \quad \times \left[
\left(1+\frac{3T_0}{2}\frac{dc_{s0}^2}{dT_0}\right)N_L(t,\bm k)+ 2N_T(t,\bm k)
\right]. \nonumber 
\end{align}
This integral is divergent and is regularized by introducing a cut off $\Lambda$ for $K$ (not $k$).
Substituting the solution \eqref{eq:kinsol}, we write the fluctuating contribution as a sum of two terms,
\st
\label{eq:Tii_fluctdef}
  T^{ii}_{\rm fluct}(t,\Lambda) = T^{ii}_{N_{\rm eq}}(t;\Lambda)  + 
  T^{ii}_{\delta N}(t;\Lambda).
\stp
The first term arises from equilibrium fluctuations $N_{\rm eq}(t)$ (\Eq{eq:kinsol_eq}) 
\begin{align}
\label{eq:Tii_cubic}
T^{ii}_{N_{\rm eq}}(t;\Lambda)
\equiv [1-h(t)] \left(1+\frac{T_0}{2}\frac{dc_{s0}^2}{dT_0}\right) \frac{T_0\Lambda^3}{2\pi^2}, 
\end{align}
while the second term  arises from the nonequilibrium distribution functions, $\delta N_{L/T}$ in \Eq{eq:kinsol_neq}. 
In frequency space this  nonequilibrium contribution reads
\begin{align}
\label{eq:Tii_rest}
   T^{ii}_{\delta N}(\omega;\Lambda)  \equiv &
 \frac{h(\omega)\bar T_0}{4\pi^2}\left(1+\frac{3\bar T_0}{2}\frac{d\bar c_{s0}^2}{d\bar T_0}\right) 
\bar C_{\zeta 0}f(\omega,\bar\gamma_{\zeta 0},\Lambda) \nonumber \\
& \quad +\frac{h(\omega)\bar T_0}{\pi^2}\bar C_{\eta 0}f(\omega,2\bar\gamma_{\eta 0},\Lambda).
\end{align}
Here we have defined a function,
\begin{align}
\label{eq:freg}
f(\omega,\gamma,\Lambda)
&\equiv
\int_0^{\Lambda\to\infty}p^2dp \frac{i\omega}{-i\omega + \gamma p^2} \\
&=\frac{i\omega}{\gamma}\Lambda
-\left(\frac{|\omega|}{\gamma}\right)^{3/2}\frac{\pi}{2\sqrt{2}}(1+ i {\rm sgn}(\omega)). \nonumber 
\end{align}

Next, we calculate the averaged energy density in a similar manner.
It also consists of contributions from the background fluid and from the two-point functions of the fluctuations:
\begin{subequations}
\begin{align}
   \label{eq:defTtt}
\langle T^{tt}\rangle &=e_0 + T^{tt}_{\rm fluct}, \\
T^{tt}_{\rm fluct} &=\frac{\langle \vec G^2 \rangle}{e_0+p_0} \\
&=\frac{1}{e_0+p_0}\int\frac{d^3k}{(2\pi)^3}
\left[N_L(t,\bm k) + 2N_T(t,\bm k)\right], \nonumber 
\end{align}
\end{subequations}
The contribution from the fluctuations is again divergent and we regularize with the same cut-off $\Lambda$ on $K$.
Substituting the perturbative solutions \eqref{eq:kinsol}, we find
\begin{align}
   \label{eq:defTttflucts}
   T^{tt}_{\rm fluct}(t;\Lambda) =&  T^{tt}_{N_{\rm eq}}(t;\Lambda) + T^{tt}_{\delta N}(t;\Lambda),
\end{align}
where  the first term arises from the equilibrium distribution $N_{\rm eq}$ (\Eq{eq:kinsol_eq})
\begin{align}
\label{eq:Ttt_cubic}
   T^{tt}_{N_{\rm eq}}(t;\Lambda) \equiv \frac{T_0\Lambda^3}{2\pi^2}, 
\end{align}
while the second term (in frequency space) arises from $\delta N_{L/T}$,
\begin{align}
\label{eq:Ttt_rest}
T^{tt}_{\delta N} (\omega;\Lambda)
   \equiv&  \frac{h(\omega)\bar T_0}{4\pi^2}
\bar C_{\zeta 0}
f(\omega,\bar\gamma_{\zeta 0},\Lambda) \, \nonumber \\
& \quad +\frac{h(\omega)\bar T_0}{\pi^2}\bar C_{\eta 0}
f(\omega, 2\bar\gamma_{\eta 0},\Lambda).
\end{align}

As will be described in the next section, the divergences in $ T^{ii}_\text{fluct}$  and $T^{tt}_\text{fluct}$ are absorbed by renormalizing the background fields, e.g.,\ $p_0$ and $\zeta_0$.
This renormalization procedure requires a clearer understanding how these bare parameters are defined, and how they depend on the cut-off $\Lambda$. 

\subsection{Temperature shift\label{sec:tempshift}}

The bare parameters $e_0, p_0, T_0, \zeta_0, \ldots$ are determined by modes (such as particlelike excitations) with wave numbers above the cut-off,  $k > \Lambda$, which are not explicitly propagated by the statistical hydrodynamic system.
The goal of this section is to carefully explain how these parameters are defined and related to the physical equation of
state $e(T),\,p(T)$ (from lattice QCD) and the cut-off $\Lambda$. 

First consider the density matrix for nonhydrodynamic modes with wave numbers above the cut-off $k > \Lambda$.
When the system is driven slightly out of equilibrium by the periodic compression and expansion, the density matrix for these modes $\rho(\Lambda)$ can be decomposed as an equilibrium density matrix $\rho_{\rm eq}(T_0;\Lambda)$ and a nonequilibrium correction which is well characterized by a single gradient $\delta \rho_{\rm neq}(\Lambda)\propto \nabla \cdot u$,
\st
 \rho(\Lambda) = \rho_{\rm eq}(T_0;\Lambda) + \delta \rho_{\rm neq}(\Lambda) \, .
\stp
The temperature parameter $T_0$ (which will depend on time and $\Lambda$) is chosen so that the average energy density above the cut-off $e_0(t,\Lambda) \equiv \llangle T^{tt}(t)\rrangle_{k>\Lambda}$ equals the energy from the equilibrium density matrix $ \rho_{\rm eq}(T_0;\Lambda)$ alone
\st
  \label{eq:landau_constraint}
  e_0(t,\Lambda) \equiv \llangle T^{tt}(t)\rrangle_{k>\Lambda} = e_{\rm eq,0}(T_0(t;\Lambda); \Lambda) \, ,
\stp
i.e., $T_0(t;\Lambda)$ is adjusted so that the energy moment associated with $\delta  \rho_{\rm neq}(\Lambda)$ is zero $\delta e_{\rm neq}(t;\Lambda)=0$. 
(Otherwise the rhs of \Eq{eq:landau_constraint} would have a correction proportional to $\nabla \cdot u$).
Because of the constraint in \Eq{eq:landau_constraint} we can drop the ``eq" label below, i.e.,
\st
e_0(t,\Lambda) = e_{\rm eq,0}(T_0(t;\Lambda);\Lambda)=e_0(T_0(t;\Lambda);\Lambda). 
\stp
In kinetic theory a similar constraint is imposed by requiring that the viscous correction to the distribution function $\delta f_{\rm bulk}(\p)$ does not change the energy in the system~\cite{Arnold:2006fz,Moore:2008ws}.
Once this prescription for $T_0(t;\Lambda)$ is adopted,
the stress  computed with the density matrix $\rho(\Lambda)$ is given by\footnote{In the current setup $\nabla \cdot u=\tfrac{3}{2} \dot h$.} 
\st
   \llangle T^{ij} \rrangle_{k>\Lambda} =  (1- h) \, p_0(T_0;\Lambda) \, \delta^{ij} -  \zeta_0(T_0;\Lambda) \,  \nabla \cdot u \, \delta^{ij} \, , 
\stp
where the partial pressure $p_0(T_0;\Lambda)$ from modes above $\Lambda$ is determined by the equilibrium density matrix, $\rho_{\rm eq}(T_0;\Lambda)$, while the bulk term comes from the viscous correction, $\delta \rho_{\rm neq}(\Lambda)$. 
This is the parametrization of the stress tensor (for $k>\Lambda$) that was used in \Eq{eq:hydro+noise}.
The spatial stress tensor
determines the bulk viscous correction $\zeta_0 \nabla \cdot u$ 
only after the parameter $T_0(t;\Lambda)$ is defined according to the Landau constraint in \Eq{eq:landau_constraint}~\cite{Arnold:2006fz,Moore:2008ws}. 

Later in this section we will define a temperature $T(t)$ by imposing the Landau constraint on the whole system (including the energy of hydrodynamic fluctuations below the cut-off),  and this will lead to a difference between $T_0(t;\Lambda)$ and the cutoff independent temperature $T(t)$.

Now we will relate the partial energy density and pressure, $e_{0}(T_0;\Lambda)$ and $p_{0}(T_0;\Lambda)$, to the equilibrium energy density and pressure, $e(T_0)$ and $p(T_0)$,  as measured by lattice QCD.
Indeed, $e_0$ and $p_0$ are cut-off dependent quantities and are determined by an equilibrium density matrix $\rho_{\rm eq}(T_0;\Lambda)$ which excludes equilibrium hydrodynamic fluctuations below the scale $\Lambda$. 
The contribution of such equilibrium hydrodynamic fluctuations to the energy density and pressure are given by \Eq{eq:Ttt_cubic} and \Eq{eq:Tii_cubic}, respectively, and thus the physical energy density and pressure are: 
\begin{subequations}
\begin{align}
   \label{eq:erenorm}
   e(T_0) =&  e_0(T_0;\Lambda) + \frac{T_0 \Lambda^3}{2\pi^2}\, , \\
   \label{eq:prenorm}
   p(T_0) =&  p_0(T_0;\Lambda) + \left(1 + \frac{T_0}{2} \frac{dc_{s}^2}{dT_0}\right) \frac{T_0 \Lambda^3}{6\pi^2}  \, .
\end{align}
\end{subequations}
At a practical level these equations serve to define the $e_0$ and $p_0$  parameters that should be used in a stochastic hydrocode with a given  cut-off $\Lambda$ and physical equation of state $e(T_0),\,p(T_0)$.

As discussed above, the temperature $T(t)$ for the complete system (background+fluctuations) is adjusted so that the energy density calculated from the lattice equation of state $e(T(t))$ matches the energy of the partially equilibrated system $\llangle T^{tt}(t) \rrangle$, 
\st
  \llangle T^{tt}(t) \rrangle = e(T(t))  \, .
\stp
After imposing this constraint, the time-dependent stress $\llangle T^{ii}(t) \rrangle$ of the driven system will deviate from its equilibrium expectation, $3\, p(T(t)) \, (1- h(t))$, and these deviations are described (up to long-time tails) by the bulk viscosity.
Combining Eqs.~(\ref{eq:defTtt}), (\ref{eq:defTttflucts}), (\ref{eq:Ttt_cubic}), and (\ref{eq:erenorm}), the energy of the background+fluctuations is
\begin{align}
   \label{eq:DTdeff}
   e(T(t)) &= e(T_0(t;\Lambda)) +  T^{tt}_{\delta N}(t;\Lambda),
\end{align}
where $T^{tt}_{\delta N}(t;\Lambda)$ was defined in \Eq{eq:Ttt_rest}.
Thus, the temperature for the whole system $T(t)$ (which is
independent of the cut-off) is related to the temperature
parameter of the subsystem $T_0(t;\Lambda)$ by a small shift $\Delta T$
\st
 \label{eq:Tshiftdef}
 T_{0}(t;\Lambda) = T(t) + \Delta T(t;\Lambda) \, ,
\stp
so that \Eq{eq:DTdeff} is satisfied.
The temperature shift is given in frequency space by
\begin{align}
\label{eq:T_shift}
   -\frac{de}{dT}\Delta T(\omega;\Lambda)
&=\frac{h(\omega)\bar T}{4\pi^2}
\bar C_{\zeta 0}f(\omega,\bar\gamma_{\zeta 0},\Lambda) \\
& \quad +\frac{h(\omega)\bar T}{\pi^2}\bar C_{\eta 0}f(\omega, 2\bar\gamma_{\eta 0},\Lambda). \nonumber
\end{align}
and clearly depends on the cutoff because the $T_0(t;\Lambda)$ was defined with respect to a specific subsystem labeled by $\Lambda$.
The temperature shift in the time domain takes the form 
  \begin{multline}
 \label{eq:Tshiftdiverge}
  -\frac{de}{dT}  \Delta T(t;\Lambda) = -\frac{\bar T\Lambda}{6\pi^2} \left[\frac{\bar C_{\zeta 0} }{\bar\gamma_{\zeta0}} +  4\frac{\bar C_{\eta 0} }{2\bar \gamma_{\eta 0}}\right] \nabla \cdot u \\ + {\rm finite} \, ,
  \end{multline}
where $\nabla \cdot u = \tfrac{3}{2} \dot h$ for this example.
The divergent piece of the temperature shift is universal, but the finite corrections are not.
This is verified by explicit calculation of the temperature shift for the Bjorken background in \app{app:bjorken}.
From practical perspective,  \Eqs{eq:Tshiftdef} and (\ref{eq:Tshiftdiverge}) define how $T_0$ must be chosen for a stochastic hydro code (with a specified cut-off $\Lambda$) to  reproduce the correct physical bulk viscosity for long wavelength hydrodynamic modes and a physical equation of state.
This is detailed in the next section\footnote{In defining $T_0$ from $T$, $\Lambda$, and $\nabla \cdot u$, the finite remainder in \Eq{eq:Tshiftdiverge} can be chosen in any convenient way.}.

\subsection{Renormalized background and long-time tails}

Once the temperature shift $\Delta T(t;\Lambda)$ is obtained, the remaining divergences in $T^{ii}_{\rm fluct}$ can be absorbed by pressure and bulk viscosity renormalization.
Using Eqs.~(\ref{Tijtotdef}), (\ref{eq:Tii_fluctdef}), (\ref{eq:Tii_cubic}), and (\ref{eq:prenorm}), the statistically averaged spatial stress tensor trace reads
\begin{multline}
   \langle T^{ii}\rangle(t)=3\left[1-h(t)\right] p(T_0(t;\Lambda)) \\
 -\frac{9}{2}\dot h(t)\zeta_0 + T^{ii}_{\delta N}(t;\Lambda).
\end{multline}
where $T^{ii}_{\delta N}$ is given in \Eq{eq:Tii_rest}.

Now we will shift the temperature parameter $T_0$ in the pressure to the physical temperature $T(t)$ determined by Landau matching \eq{eq:DTdeff}, $p(T_0) = p(T) + p'(T) \Delta T$.
The fluctuation contribution $T^{ii}_{\delta N}(t;\Lambda)$ and the temperature parameter $\Delta T(t;\Lambda)$ both diverge as $-i\omega h(\omega)\Lambda$.
These two terms gracefully
combine to produce a positive definite renormalization of bulk viscosity $\zeta_0$ in the term $- \frac{9}{2}\dot h(t)\zeta$,
\begin{align}
\label{eq:bulk_ren}
\zeta(T)=\zeta_0(T;\Lambda)+\frac{T\Lambda}{18\pi^2}
\left[ 
\frac{C_{\zeta 0}^2}{\gamma_{\zeta 0}}+4\frac{C_{\eta 0}^2}{2\gamma_{\eta 0}}
\right] \, .
\end{align}
In this step the coefficients in front of the linear divergences in $\Delta T$ and $T^{ii}_{\delta N}$ have neatly come together to complete the squares of $C_{\zeta 0}$ and $C_{\eta 0}$  defined by \Eq{eq:C}. 
Thus the renormalization of the bulk viscosity is positive and only necessary in a system with broken scale symmetry.
We have confirmed that the bulk viscosity renormalization is universal by computing it for a Bjorken expanding background (see Appendix \ref{app:bjorken}).

Once all divergences are absorbed by renormalization, the stress tensor becomes finite and cut-off independent.
In the presence of background expansion, there are remaining finite corrections from the fluctuations in $T^{ii}_{\delta N}$.
The total stress tensor is 
\begin{align}
\langle T^{ii}\rangle(t)
&=3\left[1-h(t)\right] p(T(t))
-\frac{9}{2}\dot h(t)\zeta(T(t)) \nonumber \\
& \quad -\int \frac{d\omega}{2\pi}e^{-i\omega t}h(\omega)
|\omega|^{3/2} \frac{\pi}{2\sqrt{2}}(1+i{\rm sgn}(\omega)) \nonumber \\
& \quad\quad \times \frac{\bar T}{4\pi^2}\left[
\bar C_{\zeta 0}^2 \left(\frac{1}{\bar \gamma_{\zeta 0}}\right)^{3/2}
+4\bar C_{\eta 0}^2\left(\frac{1}{2\bar\gamma_{\eta 0}}\right)^{3/2}
\right]\nonumber ,\label{eq:finite}
\end{align}
and has a term with $|\omega|^{3/2}$, which cannot be expressed by local time derivatives.
This term is not analytic at $\omega=0$ and derives from the out-of-equilibrium fluctuations in the kinetic regime $k\sim k_*$. 

With $\langle T^{tt}\rangle$ and $\langle T^{ii}\rangle$ known, we can write down the hydrodynamic equations for statistically averaged hydrodynamics with noise
\begin{align}
0=\frac{d}{dt}\langle T^{tt}\rangle + \frac{3}{2}\dot h \langle T^{tt}\rangle + \frac{1}{2}\dot h \langle T^{ii}\rangle .
\end{align}
Because the nonanalytic term in $\langle T^{ii}\rangle$ is of $\mathcal O(h)$, the rest frame energy density $e(t)$ evolves according to
\begin{align}
0=\dot e(t) + \frac{3\dot h}{2}\left[e(t) + p(t)\right] ,
\end{align}
and we obtain the solution: 
\begin{align}
e(t) = \bar e - \frac{3h(t)}{2}(\bar e + \bar p),
\end{align}
which will be used to calculate the response function in the next section.

\subsection{Response function in the bulk channel\label{sec:GR}}
The nonanalytic behavior in $\omega$ is also present in the response function in the bulk channel.
In the frequency space, the linear response of stress tensor to the external gravitational field $h(\omega)$ is given by
\begin{align}
\langle T^{ii}\rangle(\omega)
=G_R^{ii,jj}(\omega,\bm k=\bm 0)\frac{1}{2}h(\omega).
\end{align}
The response function $G_R^{ii,jj}$ is defined by
\begin{subequations}
\label{eq:bulk_retarded}
\begin{align}
G_R^{ii,jj}(t,\bm x) &\equiv i\theta(t)
\left\langle\left[ \hat T^{ii}(t,\bm x),\hat T^{jj}(0, \bm 0) \right]\right\rangle, \\
G_R^{ii,jj}(\omega,\bm k)&=\int d^4 x\,G_R^{ii,jj}(t,\bm x) e^{i\omega t -i\bm k\cdot \bm x}.
\end{align}
\end{subequations}
Then from our results, the response function $G_R^{ii,jj}(\omega)\equiv G_R^{ii,jj}(\omega,\bm k = \bm 0)$ is obtained as
\begin{align}
& G_R^{ii,jj}(\omega)
=\frac{\delta}{\delta h(\omega)}\left[2 \langle T^{ii}\rangle(\omega)\right] \Big |_{h=0}\nonumber \\
&= -6\left(\bar p+\frac{3}{2}\bar c_s^2(\bar e + \bar p)\right) 
+ 9i\omega \bar \zeta  -\frac{1+i{\rm sgn}(\omega)}{4\sqrt{2}\pi} |\omega|^{3/2}\bar T \nonumber \\
& \quad \times \left[
\bar C_{\zeta}^2\left(\frac{1}{\bar \gamma_{\zeta}}\right)^{3/2}
+4\bar C_{\eta}^2\left(\frac{1}{2\bar\gamma_{\eta}}\right)^{3/2}
\right],\label{eq:GRiijj}
\end{align}
and the spectral function as
\begin{align}
&\rho^{ii,jj}(\omega)=2{\rm Im}\, G_R^{ii,jj}(\omega)\\
&=18\omega \bar \zeta
-\frac{\omega|\omega|^{1/2} \bar T}{2\sqrt{2}\pi}
\left[
\bar C_{\zeta}^2\left(\frac{1}{\bar\gamma_{\zeta}}\right)^{3/2}
+4\bar C_{\eta}^2\left(\frac{1}{2\bar\gamma_{\eta}}\right)^{3/2}
\right]. \nonumber
\end{align}
This spectral function is consistent with a previous diagrammatic computation of the symmetrized correlation function $C^{ii,jj}$ (see the appendix of \Ref{Kovtun:2003vj}) using the fluctuation-dissipation relation:\footnote{
The term $18\omega\bar\zeta$ in $\rho^{ii,jj}(\omega)$ corresponds to a correlation of thermal noise in the stress tensor, which is not explicitly written in the calculation of $C^{ii,jj}$ \cite{Kovtun:2003vj}.
}
\begin{subequations}
\begin{align}
\rho^{ii,jj}(\omega)&=\frac{\omega}{T} C^{ii,jj}(\omega,\bm k = \bm 0), \\
C^{ii,jj}(t,\bm x) &\equiv \frac{1}{2}
\left\langle \left\{\hat T^{ii}(t,\bm x),\hat T^{jj}(0, \bm 0) \right\}\right\rangle_{\rm conn}.
\end{align}
\end{subequations}

We also computed $G_R^{ii,jj}(\omega)$ diagrammatically in \app{app:diagrams} and found identical results to \Eqs{eq:GRiijj} and (\ref{eq:bulk_ren}) up to a contact term\footnote{
Deviation by a contact term is permitted because of different definitions of the two-point functions \cite{Romatschke:2009ng}.
}.


\section{Summary}
\label{sec:summary}

In this paper we applied the kinetic theory of hydrodynamic fluctuations developed in~\Ref{Akamatsu:2016llw} to a relativistic nonconformal fluid.
We calculated the contribution of out-of-equilibrium hydrodynamic fluctuations to the energy momentum tensor, which renormalize the background hydrodynamic fields and the bulk viscosity $\zeta$.
The bulk viscosity renormalization is proportional to the scaling symmetry breaking in the equation of state and can be used to estimate the minimal bulk viscosity value in a hot QCD medium.

In the main body of the paper, we considered a nonconformal charge-neutral fluid, which is driven out of equilibrium by a weak isotropic expansion (or compression).
Analogous calculations for a Bjorken expanding system is summarized in the appendix.
The relaxation of hydrodynamic fluctuations to equilibrium is disturbed by the expansion and the deviation of two-point correlations from equilibrium becomes appreciable for wavelengths $k\alt k_*\sim\sqrt{\omega/\gamma_{\eta,\zeta}}$, where $\omega$ is the frequency of the background expansion and $k_*$ defines the hydrokinetic regime.

We derive the hydrokinetic equations for the two-point correlation functions $N_{AA}(t,\bm k)$, \Eq{eq:NAA}, of energy $\delta e$ and momentum  $\vec{g}$ density fluctuations in the presence of the expansion.
The nonlinear fluctuations  $N_{AA}(t,\bm k)$ contribute to the statistically averaged energy-momentum tensor $\left<T^{\mu\nu}\right>$.
The divergent part of the fluctuation contributions is regulated by an ultraviolet cut-off $\Lambda$.
The cutoff dependence of $T^{\mu\nu}_{\rm fluct}$ is (partially) absorbed by a universal renormalization of the background energy density $e_0$, the pressure $p_0$, and the bulk viscosity $\zeta_0$ (the same terms are found for the far-from-equilibrium Bjorken expansion\footnote{
Because of the Bjorken expansion is anisotropic, there is an additional linear divergence which renormalizes the background shear viscosity \eqref{eq:shear-ren}:
\begin{align}
\eta(T) &=
\eta_0(T;\Lambda) 
+\frac{T\Lambda}{30\pi^2}
\left[
\frac{e_0+p_0}{\zeta_0 + \frac{4}{3}\eta_0}
+\frac{7(e_0+p_0)}{2\eta_0}
\right] . \nonumber
\end{align}
This is a generalization from the conformal case \cite{Akamatsu:2016llw, Kovtun:2011np}.
}; see Appendix \ref{app:bjorken}):
\begin{subequations}
\label{eq:renormalization}
\begin{align}
e(T) &= e_0(T;\Lambda) + \frac{T\Lambda^3}{2\pi^2},\label{eq:eequiv} \\
p(T) &= p_0(T;\Lambda) + \left(1+\frac{T}{2}\frac{dc_{s0}^2}{dT}\right) \frac{T\Lambda^3}{6\pi^2}, \\
\zeta(T)&=
\zeta_0(T;\Lambda) \label{eq:bulkfinal}\\
& \quad+\frac{T\Lambda}{18\pi^2}
\left[
\begin{aligned}
&\left(1+\frac{3T}{2}\frac{dc_{s0}^2}{dT} -3c_{s0}^2\right)^2\frac{e_0 + p_0}{\zeta_0+\frac{4}{3}\eta_0} \\
&+4\left(1-3c_{s0}^2\right)^2\frac{e_0 + p_0}{2\eta_0}
\end{aligned}
\right]. \nonumber
\end{align}
\end{subequations}
The bare unrenormalized background quantities reflect the physical properties of the modes above the cut-off $\Lambda$.
The hydrodynamic fluctuations below the cutoff are dynamical in the hydrodynamics with noise and make an evolving contribution to the energy momentum tensor.
We find that the renormalization of the bulk viscosity is proportional to the nonconformality of the equation of state, e.g.,\ $(1-3c_{s0}^2)^2$, in agreement with other estimates~\cite{Arnold:2006fz,Meyer:2007dy,Kharzeev:2007wb,Karsch:2007jc,Moore:2008ws,Romatschke:2009ng}.
Using the parametrization of the equation of state from the lattice QCD simulations, we find that the fluctuation-induced bulk viscosity is modestly enhanced around the QCD pseudocritical temperature $T_c\sim150\,\text{MeV}$, where deviations from  the conformality are the largest (see \Fig{fig:bulk})~\cite{Borsanyi:2016ksw}.
A diagrammatic derivation of similar bound for bulk viscosity for a nonrelativistic cold Fermi gas was recently presented in \Ref{Martinez:2017jjf} and we performed the calculation for the relativistic nonconformal fluid in \app{app:diagrams} confirming the bulk viscosity renormalization, \Eq{eq:bulkfinal}.

In a nonconformal system, the contribution to the energy density from the hydrodynamic fluctuations  $T^{tt}_{\rm fluct}$ is not completely accounted for by the equilibrium energy density of hydrodynamic modes (the cubic term in \Eq{eq:eequiv}). 
The additional cutoff dependent contributions are proportional to the divergence of the flow velocity $\nabla \cdot u$ and are removed by a universal shift in the background temperature $T_0=T(\Lambda)+\Delta T(\Lambda)$, \Eq{eq:Tshiftdiverge}.
Once the cutoff dependence in $T^{\mu\nu}_{\rm fluct}$ is completely absorbed, the remaining finite contribution has a fractional power in the gradient expansion ($\propto \omega^{3/2}$) and makes an essential difference from hydrodynamics without noise (see \Eq{eq:finite}).
In the symmetrized correlation function of the energy-momentum tensor $C^{ii,jj}$, these terms become proportional to $\omega^{1/2}$ and in coordinate space only decay with a power law tail $\propto t^{-3/2}$, and therefore are called the long-time tails.
Comparing the spectral functions $\rho^{ii,jj}$, we find that our computation using the hydrokinetic theory is consistent with the previous diagrammatic calculations~\cite{Kovtun:2003vj}.

In this publication we extended our previous work on hydrokinetic theory to nonconformal systems close to equilibrium and undergoing a Bjorken expansion. 
A natural next step is to consider more general background evolution and systems with the net baryon number. It would be particularly rewarding to extend the hydrokinetic theory to critical fluctuations around the critical point, which is the focus of the beam energy scan program at RHIC.

\begin{acknowledgments}
	This work was supported in part by the U.S. Department of Energy, Office of 
	Science, Office of Nuclear Physics  under Award Number 
	DE\nobreakdash-FG02\nobreakdash-88ER40388 (A.M., D.T.). 
	This work was also supported in part by the German Research Foundation (DFG) 
	Collaborative Research Centre (SFB) 1225 Isolated quantum systems and universality in extreme conditions (ISOQUANT) (A.M.). 
	Y.A. thanks the DFG Collaborative Research Centre 1225 (ISOQUANT) for hospitality during his stay at Heidelberg University.
	\end{acknowledgments}

\appendix

\section{Bjorken background}\label{app:bjorken}
In this section we generalize the hydrokinetic equations for Bjorken expansion~\cite{Akamatsu:2016llw} to a nonconformal fluid.
In the case of a Bjorken expansion, the space-time metric of a comoving frame is given by
\begin{align}
ds^2=-d\tau^2 + dx^2 + dy^2 + \tau^2d\eta^2,
\end{align}
on which a background solution satisfies
\begin{align}
\frac{de_0}{d\tau} = -\frac{e_0 + p_0}{\tau} \left[1 - \frac{\gamma_{\zeta 0}}{\tau} + \cdots\right] ,
\end{align}
where on the right-hand side we keep only the first-order term in the hydrodynamic gradient expansion.
The evolution of the fluctuations $e=e_0+\delta e$, $\vec g\equiv (e_0+p_0)\vec v$ is concisely expressed by introducing the vielbein variables,
\begin{subequations}
\begin{align}
\vec G &= (G_{\hat x},G_{\hat y},G_{\hat z}) \equiv (g^x, g^y, \tau g^{\eta}), \\
\vec K &= (K_{\hat x},K_{\hat y},K_{\hat z}) \equiv (k_x, k_y, k_{\eta}/\tau),\\
\vec \Xi &= (\Xi_{\hat x},\Xi_{\hat y},\Xi_{\hat z}) \equiv (\xi^x, \xi^y, \tau \xi^{\eta}),
\end{align}
\end{subequations}
with which we define $\phi_a \equiv (c_{s0}\delta e, \vec G)$.
The evolution equation for $\phi_a$ is of the same form with the weak metric perturbation \Eq{eq:phi_evol}:
\begin{subequations}
\begin{align}
&-\dot\phi_a(\tau,\bm k)
=i\mathcal{L}_{ab}\phi_b + \mathcal{D}_{ab}\phi_b + \Xi_a + \mathcal{P}_{ab}\phi_b,\\
&\langle \Xi_a(\tau,\bm k)\Xi_b(\tau', -\bm k') \rangle
=2\mathcal{D}_{ab}\frac{T_0(e_0+p_0)}{\tau} \nonumber \\
& \qquad\qquad\qquad\qquad 
\times (2\pi)^3\delta(\bm k-\bm k')\delta (\tau-\tau'),
\end{align}
\end{subequations}
with $\mathcal L$ and $\mathcal D$ given by Eqs.~\eqref{eq:matrixL} and \eqref{eq:matrixD}.
The coupling to the background $\mathcal P$ takes a form specific to the Bjorken flow:
\begin{align}
&\mathcal{P} = \frac{1}{\tau}\left(
\begin{array}{cccc}
1+c_{s0}^2 + \frac{T_0}{2}\frac{d c_{s0}^2}{dT_0} & & &\\
 & 1 & & \\
 & & 1 & \\
 & & & 2
\end{array}
\right).
\end{align}
The four modes of the fluctuations $\phi_A\equiv \phi_a(e_A)_a \ (A=+,-,T_1,T_2)$ are defined using $e_A$'s in Eq.~\eqref{eq:vectors}, the eigenvectors of $\mathcal L$.
They are given in the polar coordinates by the following real orthonormal vectors:
\begin{subequations}
\begin{align}
\hat K &\equiv (\sin\theta_K\cos\varphi_K, \sin\theta_K\sin\varphi_K, \cos\theta_K), \\
\vec T_1 &\equiv (-\sin\varphi_K,\cos\varphi_K,0),\\
\vec T_2 &\equiv (\cos\theta_K\cos\varphi_K, \cos\theta_K\sin\varphi_K,-\sin\theta_K).
\end{align}
\end{subequations}
The evolution of the two-point functions \Eq{eq:2pt-func} is given by
\begin{subequations}
\begin{align}
	\label{eq:Bj++}
\frac{\partial}{\partial \tau}N_{\pm\pm}
=& -\gamma_{\zeta 0}K^2
\left[N_{\pm\pm} - \frac{T_0(e_0+p_0)}{\tau} \right] \\
&-\frac{1}{\tau}\left[2+c_{s0}^2+ \frac{T_0}{2}\frac{d c_{s0}^2}{dT_0}+\cos^2\theta_K \right]N_{\pm\pm},\nonumber\\
\frac{\partial}{\partial \tau}N_{T_1T_1}
=& -2\gamma_{\eta 0}K^2
\left[N_{T_1T_1} - \frac{T_0(e_0+p_0)}{\tau}\right]  \\
&-\frac{2}{\tau}N_{T_1T_1},\nonumber\\
\frac{\partial}{\partial \tau}N_{T_2T_2}
=& -2\gamma_{\eta 0}K^2
\left[N_{T_2T_2} - \frac{T_0(e_0+p_0)}{\tau}\right] \\
&-\frac{2}{\tau}\left[1+\sin^2\theta_K\right]N_{T_2T_2}.\nonumber
\end{align}
\end{subequations}
The only difference from a conformal case~\cite{Akamatsu:2016llw} is a term $\propto dc_{s0}^2/dT_0$ in \Eq{eq:Bj++}.
The solutions at large $K$ behave asymptotically as
\begin{subequations}
\begin{align}
\frac{N_{\pm\pm}}{T_0(e_0+p_0)/\tau} &= 1 + \frac{c_{s0}^2 -\frac{T_0}{2}\frac{dc_{s0}^2}{dT_0} - \cos^2\theta_K}{\gamma_{\zeta 0}K^2\tau} +\cdots,\nonumber\\
\ \\
\frac{N_{T_1T_1}}{T_0(e_0+p_0)/\tau} &= 1 + \frac{c_{s0}^2}{\gamma_{\eta 0}K^2\tau} +\cdots,\\
\frac{N_{T_2T_2}}{T_0(e_0+p_0)/\tau} &= 1 + \frac{c_{s0}^2-\sin^2\theta_K}{\gamma_{\eta 0}K^2\tau} +\cdots.
\end{align}
\end{subequations}

The total energy-momentum tensor is calculated from two contributions: the background part and the fluctuation part,
\begin{subequations}
\begin{align}
\label{eq:Ttt_tot}
\langle T^{\tau\tau}\rangle
&= e_0 +T^{\tau\tau}_{\rm fluct}, \\
\label{eq:Txx_tot}
\langle T^{xx}\rangle
&= p_0 - \frac{1}{\tau}\left(\zeta_0-\frac{2\eta_0}{3}\right) + T^{xx}_{\rm fluct},\\
\label{eq:Tyy_tot}
\langle T^{yy}\rangle
&= p_0 - \frac{1}{\tau}\left(\zeta_0-\frac{2\eta_0}{3}\right) + T^{yy}_{\rm fluct},\\
\label{eq:Tzz_tot}
\langle \tau^2 T^{\eta\eta}\rangle
&= p_0 - \frac{1}{\tau}\left(\zeta_0 + \frac{4\eta_0}{3}\right) + \tau^2T^{\eta\eta}_{\rm fluct},
\end{align}
\end{subequations}
with
\begin{subequations}
\begin{align}
\label{eq:Ttt_fluct}
T^{\tau\tau}_{\rm fluct}
&= \frac{\langle \vec G^2\rangle}{e_0+p_0}, \\
\label{eq:Txx_fluct}
T^{xx}_{\rm fluct}
&=\frac{\langle (G_{\hat x})^2\rangle + \frac{T_0}{2}\frac{d c_{s0}^2}{d T_0}\langle (c_{s0}\delta e)^2\rangle}
{e_0+p_0},\\
\label{eq:Tyy_fluct}
T^{yy}_{\rm fluct}
&=\frac{\langle (G_{\hat y})^2\rangle + \frac{T_0}{2}\frac{d c_{s0}^2}{d T_0}\langle (c_{s0}\delta e)^2\rangle}
{e_0+p_0}, \\
\label{eq:Tzz_fluct}
\tau^2 T^{\eta\eta}_{\rm fluct}
&= \frac{\langle (G_{\hat z})^2\rangle + \frac{T_0}{2}\frac{d c_{s0}^2}{d T_0}\langle (c_{s0}\delta e)^2\rangle}
{e_0+p_0}.
\end{align}
\end{subequations}
The $K$-space integrals are ultraviolet divergent and they are regularized by a cut-off at $|K|=\Lambda$.
The result is
\begin{subequations}
\begin{align}
T^{\tau\tau}_{\rm fluct} &= \frac{T_0\Lambda^3}{2\pi^2}
-\frac{T_0\Lambda}{6\pi^2\tau}\left[
\begin{aligned}
&\left(1 + \frac{3T_0}{2}\frac{d c_{s0}^2}{dT_0} - 3c_{s0}^2\right)\frac{1}{\gamma_{\zeta 0}} \\
&+4\left(1- 3c_{s0}^2\right) 
 \frac{1}{2\gamma_{\eta_0}}
\end{aligned}
\right] \nonumber \\
& \quad +\mathcal O(\Lambda^0), 
\end{align}
\begin{align}
T^{xx}_{\rm fluct} &= T^{yy}_{\rm fluct} \nonumber\\
&= \left(1+\frac{T_0}{2} \frac{d c_{s0}^2}{dT_0} \right)
\frac{T_0\Lambda^3}{6\pi^2}
\nonumber \\
&\quad - \frac{T_0\Lambda}{6\pi^2\tau}\left[
\begin{aligned}
&\frac{T_0}{2}\frac{dc_{s0}^2}{dT_0}
\frac{1}{\gamma_{\zeta 0}}
\left(1 + \frac{3T_0}{2}\frac{dc_{s0}^2}{dT_0} - 3c_{s0}^2 \right)\\
&+\frac{1}{\gamma_{\zeta 0}}
\left(\frac{1}{5} + \frac{T_0}{2}\frac{dc_{s0}^2}{dT_0} - c_{s0}^2 \right)\\
&+\frac{1}{2\gamma_{\eta 0}}
\left(\frac{2}{5} - 4c_{s0}^2 \right) 
\end{aligned}
\right] \nonumber\\
& \quad +\mathcal O(\Lambda^0) ,
\end{align}
\begin{align}
\tau^2 T^{\eta\eta}_{\rm fluct}
&= \left(1 + \frac{T_0}{2}\frac{d c_{s0}^2}{d T_0} \right)
\frac{T_0 \Lambda^3}{6\pi^2} \nonumber \\
& \quad -\frac{T_0\Lambda}{6\pi^2\tau}\left[
\begin{aligned}
&\frac{T_0}{2}\frac{d c_{s0}^2}{dT_0}
\frac{1}{\gamma_{\zeta 0}}
\left( 1 +\frac{3T_0}{2}\frac{dc_{s0}^2}{dT_0} - 3c_{s0}^2\right)\\
&\frac{1}{\gamma_{\zeta 0}}
\left(\frac{3}{5} + \frac{T_0}{2}\frac{dc_{s0}^2}{dT_0} - c_{s0}^2\right)\\
&+\frac{1}{2\gamma_{\eta 0}}
\left(\frac{16}{5} - 4c_{s0}^2\right)
\end{aligned}
\right]\nonumber \\
& \quad +\mathcal O(\Lambda^0) . 
\end{align}
\end{subequations}

The linear divergence in $T_{\rm fluct}^{\tau\tau}$ is absorbed by shifting the background temperature $T_0(\Lambda) = T + \Delta T(\Lambda)$:
\begin{align}
\frac{de_0}{dT_0}\Delta T
=\frac{T\Lambda}{6\pi^2\tau}
\left[
\begin{aligned}
&\left(1 + \frac{3T_0}{2}\frac{d c_{s0}^2}{dT_0} - 3c_{s0}^2\right)\frac{1}{\gamma_{\zeta 0}} \\
&+4\left(1- 3c_{s0}^2\right) 
 \frac{1}{2\gamma_{\eta_0}}
\end{aligned}
\right] + \mathcal O(\Lambda^0).
\end{align}
Noting  that $\nabla \cdot u=1/\tau$ for a Bjorken expansion, we see that this result 
agrees with \Eq{eq:Tshiftdiverge}, confirming that the divergent piece of
the temperature shift is universal.

With this temperature shift, the energy-momentum tensor is
\begin{subequations}
\begin{align}
&\langle T^{\tau\tau}\rangle
= e_0(T;\Lambda) + \frac{T\Lambda^3}{2\pi^2}, \\
&\frac{1}{3}\langle T^{xx} + T^{yy} + \tau^2T^{\eta\eta}\rangle \\
&\quad = p_0(T;\Lambda) +\left(1 + \frac{T}{2}\frac{d c_{s0}^2}{d T} \right)
\frac{T \Lambda^3}{6\pi^2} \nonumber \\
& \quad \quad -\frac{\zeta_0(T;\Lambda)}{\tau}
-\frac{T\Lambda}{18\pi^2\tau}
\left[
\frac{C_{\zeta 0}^2}{\gamma_{\zeta 0}}
+4\frac{C_{\eta 0}^2}{2\gamma_{\eta 0}}
\right] +\mathcal O(\Lambda^0), \nonumber\\
&\frac{1}{4}\langle T^{xx}+T^{yy}-2T^{\eta\eta}\rangle \\
&\quad=\frac{\eta_0(T;\Lambda)}{\tau} 
+\frac{T\Lambda}{30\pi^2\tau}
\left[
\frac{1}{\gamma_{\zeta 0}}
+\frac{7}{2\gamma_{\eta 0}}
\right] + \mathcal O(\Lambda^0), \nonumber
\end{align}
\end{subequations}
and energy density, pressure, and viscosities are renormalized as
\begin{subequations}
\begin{align}
e(T) &= e_0(T;\Lambda) + \frac{T\Lambda^3}{2\pi^2}, \\
p(T) &= p_0(T;\Lambda) + \left(1+\frac{T}{2}\frac{dc_{s0}^2}{dT}\right) \frac{T\Lambda^3}{6\pi^2}, \\
\zeta(T)&=
\zeta_0(T;\Lambda) \\
& \quad+\frac{T\Lambda}{18\pi^2}
\left[
\begin{aligned}
&\left(1+\frac{3T}{2}\frac{dc_{s0}^2}{dT} -3c_{s0}^2\right)^2\frac{e_0 + p_0}{\zeta_0+\frac{4}{3}\eta_0} \\
&+4\left(1-3c_{s0}^2\right)^2\frac{e_0 + p_0}{2\eta_0}
\end{aligned}
\right], \nonumber \\
\label{eq:shear-ren}
\eta(T) &=
\eta_0(T;\Lambda) 
+\frac{T\Lambda}{30\pi^2}
\left[
\frac{e_0+p_0}{\zeta_0 + \frac{4}{3}\eta_0}
+\frac{7(e_0+p_0)}{2\eta_0}
\right] .
\end{align}
\end{subequations}
By comparing with the renormalization in a weak metric perturbation Eq.~\eqref{eq:renormalization}, we can conclude that background field renormalization is also independent of background expansion.


\section{Long-time tails in diagrammatic approach}
\label{app:diagrams}

In this section we re-derive the retarded Green function for the trace of 
energy-momentum tensor, \Eq{eq:GRiijj}, which was
discussed in \Sec{sec:GR},
using a diagrammatic one-loop calculation.
This approach was pioneered in \Ref{Kovtun:2003vj} for the symmetric stress-stress correlations and applied to conformal and nonrelativistic fluids, respectively, in \Ref{Kovtun:2011np} and \Ref{Martinez:2017jjf}.

First we find the symmetrized Green functions for hydrodynamic fields using the 
equations of motion coupled to thermal noise.
For a static fluid, the linearized equations of motion can be Fourier transformed in frequency space from \Eq{eq:eom1} to
\begin{subequations}
	\label{eq:wkeom}
\begin{align}
&-i\omega \delta w +ic_s k^i v_i=0,\\
&-i\omega v^i +i c_s k^i \delta w\nonumber\\
&\quad+\gamma_\eta k^2 (\delta^i_{j}-\hat{k}^i\hat{k}_j)v^j+\gamma_\zeta k^2 
\hat{k}_i\hat{k}_j v^j+\tilde{\xi}^i=0,\\
&\left<\tilde{\xi}_i(\omega,\k)\tilde{\xi}_j(-\omega',-\k')\right>=\frac{2 T}{e+p}(2\pi)^4\delta(\omega-\omega')\delta(\k-\k')\nonumber\\
&
\qquad\qquad\qquad\times[\gamma_\eta
 k^2 (\delta_{ij}-\hat{k}_i\hat{k}_j)+\gamma_\zeta k^2 \hat{k}_i\hat{k}_j],
\end{align}
\end{subequations}
where for simplicity we normalize perturbations and noise by enthalpy:
\begin{align}
\delta w(\omega,\k) &= \frac{c_s\delta e(\omega,\k)}{e+p},\quad
v^i(\omega,\k) =\frac{g^i(\omega,\k)}{e+p},\nonumber\\
\tilde{\xi}^i(\omega,\k) &=\frac{\xi^i(\omega,\k)}{e+p}.
\end{align}

The symmetrized correlation function, i.e.\ the symmetrized Green function, is 
then defined as
\begin{align}
G^{\delta w, \delta w}_S(\omega,\k)= \int \frac{d\omega'}{2\pi} 
\frac{d^3\k'}{(2\pi)^3} \left<\frac{1}{2}\left\{ \delta w(\omega,\k), \delta 
w(-\omega',-\k')\right\}\right> .
\end{align}
Using the equations of motion for perturbations and the variance of noise, 
\Eq{eq:wkeom}, one easily obtains the symmetrized correlator 
between 
different combinations of hydrodynamic fields:
\begin{subequations}
\begin{align}
G^{\delta w, \delta w}_S(\omega,\k)&= \frac{2T}{e+p}   c_s^2 k^2
D^\text{sound}_S,\\
G^{v^i,v^j}_S(\omega,\k)&= \frac{2T}{e+p}  \omega^2
\left[(\delta^{ij}-\hat{k}^i\hat{k}^j)D^\text{shear}_S+\hat{k}^i\hat{k}^j 
D^\text{sound}_S \right],\\
G^{v^i,\delta w}_S(\omega,\k)&=G^{\delta w,v^i}_S(\omega,\k)
= \frac{2T  }{e+p}  c_s k^i \omega D^\text{sound}_S,
\end{align}
\end{subequations}
where common terms are given by
\begin{subequations}
\begin{align}
D_S^\text{shear}&=\frac{\gamma_\eta k^2}{\omega^4+(\gamma_\eta k^2\omega 
	)^2},\\
D_S^\text{sound}&=\frac{\gamma_\zeta 
	k^2}{(\omega^2-c_s^2k^2)^2+(\gamma_\zeta k^2\omega )^2}.
\end{align}
\end{subequations}
The retarded and symmetrized Green functions satisfy the classical dissipation-fluctuation theorem~\cite{LandauStatPart1}, 
\begin{equation}
G_S(\omega,\k)= \frac{2T}{\omega}\text{Im}\, G_R(\omega,\k),
\end{equation}
and we find the retarded Green functions by contour integration according to 
Kramers--Kronig relations~\cite{LandauStatPart1},\footnote{
			In general  the Kramers-Kronig relation holds only up to 
			subtractions of the 
			ultraviolet contribution from the spectral function.
			Therefore, strictly speaking, the real part of 
			 of the retarded Green function $G_R$
			 cannot be fixed within hydrodynamic theory.
}
\begin{align}
\label{eq:Kramers-Kronig}
G_R(\omega,\k)= 
\int\frac{d\omega'}{2\pi}\frac{2\,\text{Im}\,G_R(\omega',\k)}{\omega'-\omega-i\epsilon}.
\end{align}
The retarded Green functions for hydrodynamic fields $\delta w$ and $\vec{v}$ are 
\begin{subequations}
\label{eq:RTD_propagator}
\begin{align}
G^{\delta w, \delta w}_R(\omega,\k) &= \frac{-c_s^2 k^2}{e+p}
D^\text{sound}_R,\\
G^{v^i,v^j}_R(\omega,\k)&= \frac{1}{e+p}  
\left[(\delta^{ij}-\hat{k}^i\hat{k}^j)D^\text{shear}_R(\gamma_\eta k^2)\right.\nonumber\\
&\left. \quad+\hat{k}^i\hat{k}^j D^\text{sound}_R (-c_s^2 k^2+i\gamma_\zeta k^2 \omega)\right],\\
G^{v^i,\delta w}_R(\omega,\k)&=G^{\delta w,v^i}_R(\omega,\k)
=  \frac{-c_s k^i \omega}{e+p}   D^\text{sound}_R,
\end{align}
\end{subequations}
with
\begin{subequations}
\begin{align}
D_R^\text{shear}&=\frac{1}{-i\omega+ \gamma_\eta k^2 },\\
D_R^\text{sound}&=\frac{1}{\omega^2-c_s^2k^2 +i\gamma_\zeta k^2\omega}.
\end{align}
\end{subequations}

Similarly to the procedure in \Ref{Kovtun:2003vj}, we expand the energy-momentum 
tensor to quadratic order in perturbations (but neglect the charge density 
fluctuations),
\begin{subequations}
\label{eq:T00Tij2nd}
\begin{align}
\frac{c_s^2T^{00}}{e+p} &= \frac{c_s^2 e}{e+p} + c_s\delta w + c_s^2 \vec v^2, \\
\frac{T^{ij}}{e+p}&= \ \delta^{ij}\left[\frac{p}{e+p}+c_s\delta w +\frac{1}{2} T\frac{dc_s^2}{dT} (\delta w)^2\right]
+ v^{i} v^{j} + \frac{S^{ij}}{e+p},
\end{align}
\end{subequations}
where $S^{ij}$ denotes the thermal noise in \Eq{eq:hydro+noise}
\footnote{
By taking averages over \Eq{eq:T00Tij2nd}, we can easily find the renormalization of energy density and pressure \Eq{eq:renormalization}.
}.
We compute correlation function for
\begin{align}
\frac{{\tilde T}^{ii}}{e+p}&\equiv \frac{T^{ii} - 3c_s^2 T^{00}}{e+p} \\
&=3\left[\frac{p-c_s^2 e}{e+p}+\frac{1}{2} T\frac{dc_s^2}{dT} (\delta w)^2\right]
+ (1-3c_s^2)\vec v^2 + \frac{S^{ii}}{e+p}, \nonumber
\end{align}
where $3c_s^2T^{00}$  term is subtracted to get rid of the sound peak 
singularity.
Because $T^{00}$ is a conserved density, the subtraction does not modify the correlation function of $T^{ii}$ at $k\to 0$ so that hereafter we refer to $\tilde T^{ii}$ as $T^{ii}$.

Then the retarded Green functions for the energy-momentum tensor \Eq{eq:bulk_retarded} is
\begin{align}
&\frac{G_R^{T^{ii},T^{jj}}(\omega,\k=0)}{9(e+p)^2} 
= \frac{i\omega \zeta}{(e+p)^2} + \left(\frac{1}{3}-c_s^2\right)^2 
G_R^{\vec{v}^2, \vec{v}^2}(\omega,\bm{0})\nonumber\\
& \qquad + \left(\frac{1}{2} T\frac{dc_s^2}{dT}\right)^2G_R^{\delta w^2, \delta 
w^2}(\omega,\bm{0}) \nonumber \\
& \qquad +2\left(\frac{1}{3}-c_s^2\right)\left(\frac{1}{2} 
T\frac{dc_s^2}{dT}\right)G_R^{\delta 
w^2, \vec{v}^2}(\omega,\bm{0})\label{eq:GRTiiTjj}.
\end{align}

To evaluate \Eq{eq:GRTiiTjj}, we need to express the Green function of 
composite 
fields
\begin{equation}
G^{a^ia^j, 
a^ka^l}_R(t,\x)=i\theta(t)\left<\left[a^ia^j(t,\x),a^ka^l(0,0)\right]\right>,
\end{equation}
 in terms of two-point functions of individual fields,
\begin{align}
	G_R^{a^i a^j, a^k a^l}&(\omega,\k=0) =\int \frac{d\omega}{2\pi}\int 
	\frac{d^3\k}{(2\pi)^3}\nonumber\\
	\Bigl[
	& \ G_S^{a^i a^k}(\omega',\k)G_R^{a^j 
	a^l}(\omega -\omega',-\k) \nonumber\\
	&+  G_S^{a^i a^l}(\omega',\k)G_R^{a^j 
	a^k}(\omega-\omega',-\k)\nonumber\\
	&+  G_R^{a^i a^k}(\omega',\k)G_S^{a^j 
	a^l}(\omega-\omega',-\k)\nonumber\\
	&+  G_R^{a^i a^l}(\omega',\k)G_S^{a^j a^k}(\omega-\omega',-\k)  
	\Bigr].\label{eq:1loop}
\end{align}

Substituting appropriate symmetric and retarded Green functions to \Eq{eq:1loop} and exploiting the reflection and translational symmetries $\k\leftrightarrow -\k$, $\omega'\leftrightarrow \omega-\omega'$,  we write 
down the integrals for the Green functions necessary for the computation of \Eq{eq:GRTiiTjj}
\footnote{Note the factor of two in front of 
shear-shear term coming from the trace of $\delta^{ij}-\hat{k}^i\hat{k}^{j}$ 
and an additional minus sign in \Eq{eq:dwdwvv} from $\k\cdot(-\k)$.}
\begin{widetext}
\begin{subequations}
\begin{align}
G_R^{\vec{v}^2, 
\vec{v}^2}(\omega,\bm{0})=\frac{8T}{(e+p)^2}\int\frac{d\omega'd^3\k}{(2\pi)^4} 
&\,2\,\omega'^2D_S^\text{shear}(\omega',\k)(\gamma_{\eta}k^2)D_R^\text{shear}(\omega-\omega',-\k)\nonumber\\
+&\omega'^2D_S^\text{sound}(\omega',\k)(-c_s^2 k^2+i\gamma_\zeta k^2 
(\omega-\omega'))D_R^\text{sound}(\omega-\omega',-\k),\label{eq:soundsound}\\
G_R^{\delta w^2,\delta 
w^2}(\omega,\bm{0})=\frac{8T}{(e+p)^2}\int\frac{d\omega'd^3\k}{(2\pi)^4}&\,c_s^2
 k^2D_S^\text{sound}(\omega',\k)(-c_s^2 
k^2)D_R^\text{sound}(\omega-\omega',-\k),\label{eq:dwdwdwdw}\\
G_R^{\delta w^2, 
\vec{v}^2}(\omega,\bm{0})=\frac{8T}{(e+p)^2}\int\frac{d\omega'd^3\k}{(2\pi)^4}&\,c_s
 k\omega'D_S^\text{sound}(\omega',\k)(c_s 
k(\omega-\omega'))D_R^\text{sound}(\omega-\omega',-\k)\label{eq:dwdwvv}.
\end{align}
\end{subequations}
\end{widetext}
Note that by causality a retarded Green function $G_R(\omega,\k)$ can have 
poles only in the lower $\omega$-complex plane, so $G_R(\omega-\omega',\k)$ is 
analytic in the lower  $\omega '$-complex plane.
Therefore we will close the $\omega'$ integral in the \emph{lower complex plane of $\omega'$}, where only poles from the symmetric Green functions contribute.

For the shear-shear term in \Eq{eq:soundsound}, the symmetric Green function part can be expanded into 
\begin{align}
\omega'^2 D_S^\text{shear}(\omega',\k)
&=\frac{{i}/{2}}{\omega'+i \gamma_\eta k^2 
}-\frac{{i}/{2}}{\omega'-i\gamma_\eta k^2 },
\end{align}
where the second term does not contribute to the contour integral in the lower complex plane.
Evaluating the residue at $\omega'=-i\gamma_\eta k^2$ pole we get the shear-shear contribution
\begin{align}
&\left[G_{R}^{\vec{v}^2,\vec{v}^2}\right]^{\text{shear-shear}}(\omega,\bm{0})\nonumber \\
& \quad =\frac{8T}{(e+p)^2}\int\frac{d^3\k}{(2\pi)^3}\frac{1}{2}\frac{2\gamma_\eta
	 k^2}{-i\omega+2\gamma_\eta k^2}.
\end{align}
and the UV regulated $k<\Lambda$ integral can be straightforwardly expressed in a cubic divergent piece $\Lambda^3$ and $f(\omega, 2\gamma_\eta,\Lambda)$ defined in \Eq{eq:freg}.

The symmetric sound propagator  in $G_R^{\vec{v}^2,\vec{v}^2}$ can be also written as a sum of two terms
\begin{align}
&\omega'^2 D_S^\text{sound}(\omega',\k) \\
 &=\frac{ i\omega'/2}{\omega'^2-c_s^2k^2+i\gamma_\zeta 
	k^2\omega' }-\frac{i \omega'/2 }{\omega'^2-c_s^2k^2-i \gamma_\zeta 
	k^2\omega'}, \nonumber
\end{align}
where the second term vanishes under contour integration. The remainder can be 
further expanded as 
\begin{align}
&\frac{ i\omega'/2}{\omega'^2-c_s^2k^2+i 
	\gamma_\zeta 
	k^2\omega' }\nonumber\\
&=
\frac{i/2}{\omega_+-\omega_-}
\frac{\omega_+}{\omega'-\omega_+}
 -
 \frac{i/2}{\omega_+-\omega_-} \frac{\omega_-}{\omega'-\omega_-}.
\end{align}
Here $\omega_{\pm}$ are the positions of poles satisfying
\begin{align}
\omega_++\omega_-=-i\gamma_\zeta k^2,\quad
\omega_-\omega_+=-c_s^2 k^2.
\end{align}
For the ease of computation, the retarded function part in the sound-sound contribution in \Eq{eq:soundsound} can be also expressed in terms of $\omega_{\pm}$ as follows
\begin{align}
&\left(-c_s^2 k^2+i\gamma_\zeta k^2 (\omega-\omega') \right)
D_R^\text{sound}(\omega-\omega',-\k)\\
&=\frac{-\omega_+^2}{\omega_+-\omega_-}\frac{1}{\omega-\omega'-\omega_+}-
\frac{-\omega_-^2}{\omega_+-\omega_-}\frac{1}{\omega-\omega'-\omega_-}. \nonumber
\end{align}
Evaluating the $\omega'$ residues at $\omega'=\omega_{\pm}$, we obtain for the sound-sound piece of \Eq{eq:soundsound},
	\begin{align}
	&\left[G_R^{\vec{v}^2, 
	\vec{v}^2}\right]^{\text{sound-sound}}(\omega,\bm{0})\nonumber\\
&\quad=\frac{8T}{(e+p)^2}\int\frac{d^3\k}{(2\pi)^3} 
\frac{1}{2}\frac{\omega_+\omega_-}{(\omega_+-\omega_-)^2}\frac{\omega_++\omega_-}{\omega-\omega_+-\omega_-}\nonumber\\
&\qquad-\frac{1}{2} 
\frac{1}{(\omega_+-\omega_-)^2}\left[
\frac{\omega_+^3}{\omega-2\omega_+}+
\frac{\omega_-^3}{\omega-2\omega_-}\right].
\end{align}
In the kinetic approximation $c_s k\gg \gamma k^2,\omega$ this reduces to
\begin{align}
	&\left[G_R^{\vec{v}^2,\vec{v}^2}\right]^{\text{sound-sound}}(\omega,\bm{0}) \\
	&=\frac{8T}{(e+p)^2}
	\int\frac{d^3\k}{(2\pi)^3}
	\frac{1}{4}+\frac{1}{8}\frac{i\omega}{-i\omega+\gamma_\zeta 
	k^2}+\mathcal{O}\left(\frac{\omega^2, (\gamma k^2)^2}{(c_s k)^2}\right). \nonumber
\end{align}

Calculations for \Eqs{eq:dwdwdwdw} and \eq{eq:dwdwvv} proceed analogously. 
The result is
\begin{subequations}
\begin{align}
G_R^{\delta w^2, 
	\delta 
	w^2}(\omega,\bm{0})	&=\frac{8T}{(e+p)^2}\int\frac{d^3\k}{(2\pi)^3} \,
\frac{1}{4}+\frac{1}{8}\frac{i\omega}{-i\omega+\gamma_\zeta k^2},\\
G_R^{\delta w^2, 
	\vec{v}^2}(\omega,\bm{0})&=\frac{8T}{(e+p)^2}\int\frac{d^3\k}{(2\pi)^3} \,
0+\frac{1}{8}\frac{i\omega}{-i\omega+\gamma_\zeta k^2}.
\end{align}
\end{subequations}

The final combined result for the retarded Green functions, \Eq{eq:GRTiiTjj},  
is
\begin{align}
\label{eq:GR1-loop}
	&G_R^{T^{ii},T^{jj}}(\omega,{\bm 0}) 
	=9i\omega\zeta + \frac{T\Lambda^3}{2\pi^2} \left[2(1-c_s^2)^2 + 
	\frac{2}{3}\left(\frac{3}{2} 
	T\frac{dc_s^2}{dT}\right)^2\right]\nonumber\\
&\quad +\frac{T}{2\pi^2}\left[	4{C}_{\eta}^2\, f(\omega, 
2\gamma_\eta,\Lambda)+{C}_{\zeta}^2\,f(\omega, 
	\gamma_\zeta,\Lambda)\right].
\end{align}
To assure that the imaginary part of $G_R^{T^{ii},T^{jj}}$ is independent of the 
cutoff, the background bulk viscosity is renormalized as in 
\Eq{eq:renormalization}.
The cubic divergence in the real part of $G_R^{T^{ii},T^{jj}}$ does 
not 
have a corresponding counter term, but it is also not physical.
The ambiguity in the real part of the retarded propagators $G_R$ is 
because of the fact that in flat space time the retarded Green functions 
cannot be measured directly and only the imaginary part is determined through 
the symmetric correlation functions $G_S$.

\bibliography{refs}

\end{document}